# Speed-Up of DNA Melting Algorithm with Complete Nearest Neighbor Properties

Eivind Tøstesen[1], Fang Liu[1], Tor-Kristian Jenssen[2], and Eivind Hovig[1].

[1]*Department of Tumor Biology, The Norwegian Radium Hospital, N-0310 Oslo, Norway.*
[2]*PubGene AS, Forskningsveien 2A, P.O. Box 180 Vinderen, N-0319 Oslo, Norway.*

*Correspondence to*: Eivind Tøstesen (tel:+4722935392)(email:eivindto@radium.uio.no)

ABSTRACT:

We describe an optimized algorithm, which is faster and more accurate compared to previously described algorithms, for computing the statistical mechanics of denaturation of nucleic acid sequences according to the classical Poland-Scheraga type of model. Nearest neighbor thermodynamics has been included in a complete and general way, by rigorously treating nearest neighbor interactions, helix end interactions and isolated base-pairs. This avoids the simplifications of previous approaches and achieves full generality and controllability with respect to thermodynamic modeling. The algorithm computes subchain partition functions by recursion, from which various quantitative aspects of the melting process is easily derived, for example the base-pairing probability profiles. The algorithm represents an optimization with respect to algorithmic complexity of the partition function algorithm of Yeramian et al. (Biopolymers 1990, 30, 481-497): We reduce the computation time for a base-pairing probability profile from $O(N^2)$ to $O(N)$, where N is the sequence length. This speed-up comes in addition to the speed-up due to a multiexponential approximation of the loop entropy factor as introduced by Fixman and Freire and applied by Yeramian et al. The speed-up, however, is independent of the multiexponential approximation and reduces time from $O(N^3)$ to $O(N^2)$ in the exact case. A method for representing very large numbers is described, which avoids numerical overflow in the partition functions for genomic length sequences. In addition to calculating the standard base-pairing probability profiles, we propose to use the algorithm to calculate various other probabilities (loops, helices, tails) for a more direct view of the melting regions and their positions and sizes. This can provide a better understanding of the physics of denaturation and the biology of genomes.

## INTRODUCTION

There has been a revived interest in the classical problem of computing the thermal stability and the statistical physics of nucleic acids. As is often pointed out, the issue is fundamental to modern experimental methods in molecular biology such as DNA microarrays, PCR and gel electrophoresis (ref. 1). Recently, the problem has reappeared on the much more demanding scale of genomic sequences (ref. 2, 3, 4).

In this article we revisit and improve on two aspects of the problem: the algorithmic complexity and the thermodynamic parameters. While computer power in terms of speed and memory has increased since the algorithmic achievements of the '70s, so have the numbers and the lengths of known DNA sequences. Consequently, the limitation on algorithmic complexity is still the same strong concern as it was in the '70s. On the other hand, there have been advances in the area of deriving sequence-specific thermodynamic model parameters from experimental data, and a better understanding of what information such parameters represent.

Classically, there are three main methodological approaches with regards to the level of conformational complexity to modeling the thermodynamics of nucleic acids. The simplest models are the two-state models that predict the $T_m$ (midpoint of transition) and the melting curve (DSC or UV) of oligonucleotides (ref. 5). Usually the enthalpy change and the entropy change of the transition are calculated from thermodynamic parameters for nearest neighbor base-pairs (ref. 6, 7). These models normally also include concentration effects and other "external" conditions. However, they do not take internal degrees of freedom into consideration, melting and hybridization proceeds in a completely "on-off" manner.

The next level of complexity is the Poland-Scheraga type of models of helix-coil transitions in DNA. Here, a microstate of the double-stranded molecule is represented by a chain of binary units. The i'th unit specifies the state of the i'th base in the DNA sequence (from the 5'-end): either "1" for the closed (helical) state where the base is paired, or "0" for the open (coil or loop) state where the base is unpaired. These models do not consider molecular conformations in which a base is paired with other than its corresponding base in the complementary DNA strand. Melting proceeds by unzipping from the ends and by forming loops in the interior. Helix-coil transition models for DNA and $\alpha$-helices were developed during the 1950s–1960s (ref. 8).

The third level of complexity is RNA secondary structure modeling. In these models each position in the sequence can be base-paired with almost any other position in the sequence, in a specific hierarchical manner that avoids "cross-linked" base-pairs (bp's) (ref. 9). Algorithms for RNA secondary structure folding were developed from the late '70s and into the '80s based on dynamic programming algorithms for sequence alignment (ref. 9, 10).

Because the opening or closing of a bp is the basic degree of freedom in all three levels of approaches, it is common to define the corresponding base-pairing probabilities, that is, the probabilities of bp's being closed. For the DNA helix-coil transition models, base-pairing probabilities form a vector p(i), sometimes referred to as the melting profile or probability profile, while for RNA secondary structure models the base-pairing probabilities form a matrix p(i,j) (ref. 11). Accordingly, in this article we will classify DNA helix-coil transition models as one-dimensional (1D) and RNA secondary structure models as two-dimensional (2D). The two-state models for oligonucleotides are zero-dimensional, since only one base-pairing probability p, the total level of hybridization, is defined (ref. 5).

Note that the choice of complexity level for modeling is not strictly dictated by the type of nucleic acid. A 1D helix-coil model can be applied to hairpins in single-stranded RNA molecules and, vice versa, a 2D secondary structure model can be applied to double-stranded DNA (ref. 12).

The standard method in statistical mechanics is to compute partition functions, from which all statistical and physical information about the system, such as the base-pairing probabilities, can be extracted. By definition, the partition function $Q_{total}$ is the sum of statistical weights $\exp(-E_n/kT)$ summed over all possible microstates, n, of the system.

$$Q_{\text{total}} = \sum_n \exp(-E_n/kT). \tag{1}$$

The number of possible microstates grows exponentially with sequence length in both the 1D and 2D classes of models, thus prohibiting a straightforward summation of the partition function for long sequences. This problem was overcome by matrix multiplication methods and later by dynamic programming (recursion methods). The main reason for the success of these methods is that for most conformations, the molecule can be decomposed into elements that contribute independently to the free energy (additivity), resulting in a factorization of the statistical weight of that conformation. Table 1 shows a comparison of algorithmic complexities of different dynamic programming approaches for both 1D and 2D models.

The matrix multiplication method, which dates back to the study in physics of ferromagnetic Ising models in the '40s (ref. 13, 14), was introduced in the 1D helix-coil model by Zimm and Bragg (ref. 15) in 1959. During the '60s, there was an interplay between the two parallel studies of helix-coil transitions in biopolymers on the "biological" side (ref. 8) and Ising models, phase transitions and critical phenomena on the "physical" side (ref. 16). Two kinds of effects were considered in biopolymers: the sequence-dependent physical binding interactions between monomers in the chain and the entropic effects due to the size-dependent entropies of loops and coiled regions. In the context of a 1D Ising model, these effects correspond to interactions (potentials) with short and long range in space, respectively. When Poland and Scheraga considered the loop entropy effect (ref. 17, 8), they used the term "long-range" also in the biopolymers context. Long-range effects are represented by the loop entropy factor δ(j), which is a function of the loop size j. Theoretical treatments (ref. 18, 19, 20) predict this to be a power function $\delta(j) \propto j^{-\alpha}$ for large j with an exponent α between 1.5 and 2.2.

The drawback of the matrix multiplication method was the large size of matrices needed for treating the long-range effects rigorously. An alternative approach (ref. 21) was described by Poland in 1974. He defined recursion relations for base-pairing probabilities and certain conditional probabilities that can be solved by iteration. His approach skips the intermediate step of calculating the partition functions explicitly and jumps directly to the probabilities. The computing time of his algorithm is $O(N^2)$. Fixman and Freire (ref. 22) obtained an accelerated algorithm with computing time $O(N)$, by incorporating in Poland's recursion relations an approximation of the loop entropy factor as a sum of around 10 exponential functions. The Poland-Fixman-Freire (PFF) algorithm is available in various implementations (ref. 1, 23, 24).

In 1990, two articles introduced the use of dynamic programming techniques for calculating partition functions of nucleic acid models: Yeramian et al. described an algorithm for the 1D helix-coil model (ref. 25) and McCaskill described an algorithm for the 2D secondary structure model (ref. 11). Later, an ambitious treatment of excluded volume in a 2D secondary structure model was developed by Chen and Dill (ref. 26), and their algorithm combines dynamic programming and the matrix multiplication method.

Table 1 shows how computation times of different algorithms grow with sequence length for the calculation of a base-pairing profile. In general, the base-pairing probability of a base-pair "bp" is obtained by dividing Q("bp"), the partition function constrained to the subclass of microstates having the base-pair "bp", with the total (unconstrained) partition function $Q_{total}$. For each possible bp, the algorithm of Yeramian et al. (ref. 25) calculates the constrained partition function Q("bp") in a complete recursive sweep ("forward") along the sequence. Each sweep is done in time $t \in O(N)$ using a multiexponential approximation of the loop entropy factor, or in time $t \in O(N^2)$ using the exact loop entropy factor. The full 1D profile of base-pairing probabilities is therefore obtained in total time $O(t(N+1))$, that is, $O(N^2)$ for the approximation or $O(N^3)$ for the exact, by doing N+1 such sweeps. For long genomic sequences, this is much slower than the PFF algorithms, which take one "backward" recursive sweep followed by one "forward" sweep and thereby obtain the full profile in time $O(N)$ (approximation) or $O(N^2)$ (exact). As a remedy, Yeramian achieved a 20 time speed-up by only calculating the base-pairing probability at each 20th position in the sequence (ref. 2), which is possible since his constrained partition functions Q("bp") are calculated independently from each other.

McCaskill's dynamic programming (ref. 11) calculates a partition function in time $t \in O(N^3)$. This would give a computation time $O(N^{2+3})$ for a full 2D base-pairing profile, if using Yeramian et al.'s principle of a complete iterative calculation per bp. In contrast, McCaskill reduces the total calculation to order $O(N^3)$ by storing arrays of intermediate quantities during one partition function iteration, and reusing these quantities in the calculation of all the constrained partition functions Q("bp"). Some of these intermediate quantities are partition functions for subparts of the molecule.

In this article, we propose an improved partition function algorithm that overcomes the slowing down of Yeramian's melting profile calculation, as compared to the PFF algorithms. The total computation time is reduced from $O(N^2)$ to $O(N)$ for the multiexponential case, and from $O(N^3)$ to $O(N^2)$ for the exact loop entropy case. This speed-up by a factor $O(N)$ is independent of the multiexponential speed-up also by a factor $O(N)$. As Table 1 shows, our algorithm has the same algorithmic complexities, both in time and memory, as the PFF algorithms. Our speed-up is achieved following McCaskill's principle rather than Yeramian et al.'s principle: Our algorithm calculates and stores subchain partition functions in two recursive sweeps along the chain, one "forward" and one "backward", but unlike the "backward-forward" iterations of the PFF algorithm, our two recursive sweeps are done independently from one another. The algorithm then calculates the base-pairing probabilities in a third non-recursive sweep along the chain using the stored subchain partition functions. The price we pay for the speed-up is a memory usage that grows as $O(N)$, compared to the limited memory required by the Yeramian algorithm for storing partition function values. In some sense, we store information instead of recalculating it $O(N)$ times.

Both the PFF algorithms and the algorithm of Yeramian et al. assign stability factors to single units in the chain independently of other units. Nearest neighbor interactions were not accommodated explicitly, probably for simplicity reasons, although it was well-known that stacking interactions between neighboring bp's contribute to the stability. In 1981 Gotoh and Tagashira (ref. 27) made a "slight modification" of the PFF algorithms to take nearest neighbor effects into account. Instead of letting chain units correspond to bases in the sequence, they identified chain units with nearest neighbor pairs of bases, with a resulting chain length of N-1 instead of N, and a 16-letter alphabet per unit instead of 4. By fitting calculated and experimental melting curves, they were able to provide a parameter set for the 10 types of dinucleotides. Their N-1 scheme, together with their parameter set can also be used in the Yeramian algorithm to include nearest neighbor effects (ref. 2).

The N-1 scheme of Gotoh and Tagashira works reasonably well (ref. 28) for predicting macroscopic properties of the DNA molecule, such as melting curves. But the scheme is not optimal for handling positionally detailed information. The mapping of microstates in the N chain to microstates in the N-1 chain is neither injective (one-to-one) nor surjective (onto). For example, the motif ..101.. in the N-1 chain does not correspond to anything in the N chain. An interpretation of results in terms of the original N chain microstates is therefore not always straightforward.

The algorithm we propose here rigorously considers nearest neighbor interactions, without invoking any approximations such as the N-1 scheme of Gotoh and Tagashira. Nearest neighbor effects are included in a complete and general way in the model by using a format with three types of thermodynamic parameters for nearest neighbor bp's, isolated bp's and helix-ending bp's. Several sets of thermodynamic parameters in different formats have been published (ref. 6, 7), but since we employ a complete format, any published parameter set can be translated and included without loss of information. Neither the PFF nor the Yeramian algorithms possess this generality, since they include nearest neighbor properties in an approximate way and at most represent end interactions with a single parameter σ.

## METHODS

Where possible, we adopt the notation of Yeramian et al.(ref. 25) to illustrate both the similarities and the differences between their approach and the present approach. We note that we could alternatively have chosen a matrix multiplication notation that would have illustrated better the underlying multiplicativity principles of the physics. The chain units are numbered i=1,..,N, where N is the length of the DNA sequence. Each microstate n is represented by a string of 0's and 1's, and can be viewed as a binary representation of a number n between 0 and $2^N$ that identifies the microstate.

### Nearest Neighbors, Helix-Ends and Isolated Base-Pairs

The model represents nearest neighbor bp's, isolated bp's and helix-ending bp's by three arrays of temperature-dependent statistical weight factors. For each nearest neighbor pair [i-1,i] in the sequence we calculate $s^{11}(i)$, the statistical weight of 1-1 interactions for that pair. For each position i we calculate the statistical weight $s^{010}(i)$ of a 1 at that position with 0's on each side (an isolated bp). For each position i we calculate the statistical weight $s^{end}(i)$ of a 1 at that position with 0 on one side and 1 on the other, representing a helix-ending bp. An isolated bp and a helix-ending bp is defined similarly at the ends of the sequence (i=1 or N). All of these quantities are of course sequence-dependent and related to differences in free energy. For example,

$$s^{11}(i) = \exp(-\Delta G^{11}(i)/RT),$$

(2)

where $\Delta G^{11}(i) = \Delta H^{11}(i) - T\Delta S^{11}(i)$. These quantities can include an empirical, length-dependent correction for salt concentration (ref. 6, 29). We postpone to the Results section a discussion of how these quantities can be calculated using different thermodynamic parameter sets.

A helical segment, ...011...10..., of consecutive 1's with a and b being the positions of the first and the last 1, contributes a factor $s^{end}(a)s^{11}(a+1)s^{11}(a+2)\cdots s^{11}(b)s^{end}(b)$ to the statistical weight of the microstate, unless there is only one 1 in the segment contributing the factor $s^{010}(a)$. The helical segment can extend to the end of the chain (a=1 or b=N).

A loop, ...100...01..., from a to b, that is, a consecutive series of 0's bounded by 1's at positions a and b, contributes the loop entropy factor $\omega[2(b-a)]$, where the number of open units is $k = b - a - 1$. A constant factor σ ("cooperativity", "initiation" or "nucleation") could be absorbed in this ω function. A tail segment, that is, a series of 0's that extends to the end of the chain, contributes with the factor 1 by convention in the 1D helix-coil models (ref. 8).

The n=0 microstate (all zeros) corresponds to dissociated DNA strands and has a statistical weight of 1. The equilibrium constant of dissociation is represented by a factor β that is assigned to all microstates except the dissociated all-0's microstate.

### Recursion Relations Are Defined for Subchain Partition Functions

The symbol X is used to represent an unspecified state of a unit (either 0 or 1). Consider the subchain [1,i+1] of length i+1. We define $V_{10}(i+1)$ as the partition function of this subchain summed over the class of microstates XX...X10. We extend this definition to the special cases i=N and i=0: For i=N, $V_{10}(N+1)$ is the partition function of the whole chain summed over the class XX...X1 of microstates. For i=0, $V_{10}(1)$ is the partition function of unit 1 summed over the class of microstates 0. The only member of this class is the dissociated chain, so $V_{10}(1)$ is simply the statistical weight 1. To summarize, the vector elements $V_{10}(i)$, for i=1,...,N+1, are subchain partition functions for the

following classes:

$$\begin{bmatrix} 0 \\ 10 \\ X10 \\ \cdot \\ \cdot \\ XXXXX10 \\ XXXXXX1 \end{bmatrix}$$

Since the statistical weight factor of a tail of 0's is 1, the vector $V_{10}(i)$ is identical to the vector of partial partition functions V(i) defined by Yeramian et al (ref. 25). But there is a slight difference in formulation, as for our purpose it is important that the 1 at position i for $V_{10}(i+1)$ is not succeeded by a 1 at position i+1. From (ref. 25), it then follows that the total partition function of the whole chain is

$$Q_{\text{total}} = \sum_{j=1}^{N+1} V_{10}(j). \tag{3}$$

Consider again the subchain [1,i+1] with the rightmost units in state 10. The 1 at position i in this subchain is either an isolated bp in the microstate ...X010 or a helix-ending bp in the microstate ...X110, according to the state of unit i-1. Therefore the terms in $V_{10}(i+1)$ either contain the factor $s^{010}(i)$ or the factor $s^{end}(i)$, and we can write

$$V_{10}(i+1) = s^{010}(i) U_{01}(i) + s^{end}(i) U_{11}(i) \tag{4}$$

The quantities $U_{01}(i)$ and $U_{11}(i)$ are defined for i=2,...,N as follows: $U_{01}(i)$ is equal to the partition function of the subchain [1,i] summed over the class X...X01 and divided by the factor $s^{010}(i)$. And likewise, $U_{11}(i)$ is equal to the partition function of the subchain [1,i] summed over the class X...X11 and divided by the factor $s^{end}(i)$. In other words, the quantities $U_{01}(i)$ and $U_{11}(i)$ are "unfinished" subchain partition functions missing a factor for the unit i. They are useful for recursion when we extend the subchain with a unit i+1.

By extending a subchain one unit per step, we do a recursive build-up of the subchain partition function vectors $V_{10}$, $U_{01}$ and $U_{11}$. At step i of the recursive iteration we consider unit i as being 1 (closed). We then calculate our three subchain quantities characterizing this situation, $U_{01}(i)$, $U_{11}(i)$ and $V_{10}(i+1)$, using the quantities for shorter subchains (see Figure 1). For the calculation of $U_{01}(i)$, we must consider two cases: 1) unit i is the first closed unit in the chain causing strand association. In this case we multiply the factor β and the statistical weight of dissociated chains $V_{10}(1)$. Or 2), unit i is closing a loop of some size. For each loop size, we multiply the corresponding loop entropy factor and the partition function for the subchain at the other end of the loop. This gives us the following terms for $U_{01}(i)$, as illustrated in Figure 1(a):

$$U_{01}(i) = \beta V_{10}(1) + W \tag{5}$$

where

$$W = \sum_{j=2}^{i-1} V_{10}(j)\omega[2(i+1-j)]$$

(6)

The scalar product W is identical to the $W_{i-1}$ in (ref. 25). For the calculation of $U_{11}(i)$, we refer to Figure 1(b) and see that unit i is extending a helical segment, which contributes the factor $s^{11}(i)$. Unit i-1 is an end of that helical segment if unit i-2 is 0, otherwise it is not. This gives us two terms, as illustrated in Figure 1(b):

$$U_{11}(i) = s^{11}(i)\left(U_{01}(i-1)s^{end}(i-1) + U_{11}(i-1)\right).$$

(7)

For the calculation of $V_{10}(i+1)$, we simply use Equation 4, which is illustrated in Figure 1(c). Note that the vector $V_{10}$ is redundant, because the information is contained in the vectors $U_{01}$ and $U_{11}$. The vector could be omitted in an alternative formulation of the algorithm.

Iteration steps are done for i=3,...,N. To do the summation of $Q_{total}$ we add $V_{10}(i+1)$ to the sum in step i. Recursion is initialized as follows:

$V_{10}(1)=1$, $V_{10}(2) = \beta s^{010}(1)$, $U_{01}(2)=\beta$,
$U_{11}(2) = \beta s^{end}(1) s^{11}(2)$,
$V_{10}(3) = s^{010}(2) U_{01}(2) + s^{end}(2) U_{11}(2)$
and $Q_{total} = V_{10}(1) + V_{10}(2) + V_{10}(3)$.

## Multiexponential Approximation of the Loop Entropy Factor Gives a Faster Calculation

A multiexponential approximation of the loop entropy factor gives a faster calculation, as described by Fixman and Freire (ref. 22) and Yeramian et al (ref. 25). We have simply taken over this technique by defining the scalar product W to be identical to that of Yeramian et al. Their discussion therefore applies directly to our algorithm. This means that in the fast version of the algorithm we can calculate W by recursion, but in the slow version, using the exact loop entropy factor, we must do the summation of W at each iteration step. Here we outline the recursive calculation of W. The multiexponential approximation of the loop entropy factor is

$$\omega(j) = \sum_{m=1}^{I} A_m \exp(-B_m j).$$

(8)

We define two arrays of constants: $C1(m) = A_m \exp(-4B_m)$ and $C2(m) = \exp(-2B_m)$ for m=1,..,I.

In step i we obtain W as a sum $W = \sum_{m=1}^{I} W_{i-1}(m)$. The components $W_{i-1}(m)$ are obtained recursively:

$$W_{i-1}(m) = C2(m) W_{i-2}(m) + C1(m) V_{10}(i-1)$$

(9)

The recursion for $W_{i-1}(m)$ is initialized with $W_2(m) = C1(m) V_{10}(2)$. Note that we do not need the subscript on $W_{i-1}(m)$ (ref. 25).

## Forward and Backward Recursions Are Iterated Along the Chain

We have described an iterative procedure that goes forward from "left to right" in the chain for i=3,...,N. The inputs were the three arrays of sequence-dependent statistical weight factors and the outputs were the three subchain arrays $V_{10}(i)$, $U_{01}(i)$ and $U_{11}(i)$. Now consider the sequence of the complementary strand in the DNA molecule. Since the two strands are anti-parallel, this sequence begins at the opposite end. If we use this sequence as input to the iterative procedure described above, we obtain another set of subchain arrays that goes from "right to left". We label the two sets of arrays as $V_{10}^{\text{LR}}(i)$, $U_{01}^{\text{LR}}(i)$, $U_{11}^{\text{LR}}(i)$ and $V_{10}^{\text{RL}}(i)$, $U_{01}^{\text{RL}}(i)$, $U_{11}^{\text{RL}}(i)$. A bp at position i in the LR sequence will be at position $N+1-i$ in the RL sequence. Now consider the LR subchain [i-1, N] of length N+2-i. The class 01X...XX of microstates for this subchain is characterized by the subchain partition function $V_{10}^{\text{RL}}(N+2-i)$. Similarly we can interpret the vectors $U_{01}^{\text{RL}}$ and $U_{11}^{\text{RL}}$, for example, the subchain partition function $U_{01}^{\text{RL}}(N+1-i)$ characterizes the class 10X...XX of microstates for the LR subchain [i,N] of length N+1-i.

## Various Probabilities Are Calculated in the Second Part of the Algorithm

In the first part of the algorithm we do the forward LR recursion and the backward RL recursion, and the six subchain arrays are then used in the second part. In general, the probability of a class A of microstates is

$$p(\text{A}) = Q_{\text{A}}/Q_{\text{total}},$$

(10)

where $Q_{\text{A}}$ is the partition function summed over the class A of microstates. In the following, we exploit the Poland & Scheraga assumption that the only long-range interactions in this model are within loops. A closed unit only interacts with its neighbors. As a consequence, a fixed closed unit divides a chain into two nearly independent subchains and the constrained partition function $Q_{\text{A}}$ factorizes.

For the base-pairing probability $p_{\text{closed}}(i)$ we consider the class ...XX1XX.... There are four possible configurations of the two neighbors of unit i:

$$p(...\text{XX1XX}...) = p(...\text{X010X}...) + p(...\text{X011X}...) + p(...\text{X110X}...) + p(...\text{X111X}...).$$

The subchains to the left and right of unit i are characterized by $U_{01}^{\text{LR}}(i)$, $U_{11}^{\text{LR}}(i)$, $U_{01}^{\text{RL}}(N+1-i)$ and $U_{11}^{\text{RL}}(N+1-i)$ and we combine these with the "missing factor" for unit i:

$$p_{\text{closed}}(i) = \frac{U_{01}^{\text{LR}}(i)s^{010}(i)U_{01}^{\text{RL}}(N+1-i) + U_{01}^{\text{LR}}(i)s^{\text{end}}(i)U_{11}^{\text{RL}}(N+1-i) + U_{11}^{\text{LR}}(i)s^{\text{end}}(i)U_{01}^{\text{RL}}(N+1-i) + U_{11}^{\text{LR}}(i)U_{11}^{\text{RL}}(N+1-i)}{\beta Q_{\text{total}}}.$$

(11)

The denominator takes care of the overlap between LR and RL, since both of them contain the factor β. For the special cases of i=1 and i=N we simply have $p_{\text{closed}}(1) = V_{10}^{\text{RL}}(N+1)/Q_{\text{total}}$ and $p_{\text{closed}}(N) = V_{10}^{\text{LR}}(N+1)/Q_{\text{total}}$.

For the probability $p_{\text{loop}}(a,b)$ of a loop bounded by 1's at positions a and b we consider three independent segments of the chain,

$$p_{\text{loop}}(a,b) = V_{10}^{\text{LR}}(a+1)\omega[2(b-a)]V_{10}^{\text{RL}}(N+2-b)/\beta Q_{\text{total}}.$$

(12)

The probability of a tail of 0's from the right end of the chain to a 1 at position i is

$$p_{\text{right}}(i) = V_{10}^{\text{LR}}(i+1)/Q_{\text{total}}.$$

(13)

The probability of a tail of 0's from the left end of the chain to a 1 at position i is

$$p_{\text{left}}(i) = V_{10}^{\text{RL}}(N+2-i)/Q_{\text{total}}.$$

(14)

The probability of a helical segment of consecutive 1's from a to b is

$$p_{\text{helix}}(a,b) = U_{01}^{\text{LR}}(a)s^{\text{end}}(a)\left[\prod_{j=a+1}^{b} s^{11}(j)\right]s^{\text{end}}(b)U_{01}^{\text{RL}}(N+1-b)\bigg/\beta Q_{\text{total}},$$

(15)

where for completeness we can define $U_{01}^{\text{LR}}(1)=1$ and $U_{01}^{\text{RL}}(1)=1$.

In addition to these probabilities, it should be noted that also the fraction θ of bp's in the closed state can be calculated within this framework (ref. 25).

## Numerical Problem of Overflow/Underflow

The numerical problem of overflow/underflow is a major concern in calculations of partition functions by recursion. For some algorithms it is sufficient to do a rescaling of the partition functions. The set of partition function values encountered in the algorithm should all be rescaled with the same factor. Probabilities are unchanged by such a rescaling, since they are given as ratios between partition functions. But this can prevent overflow and underflow only if the range of partition function values (in orders of magnitude) is smaller than the range of numbers that can be represented in the machine. Then rescaling can "move" these values inside the machine range. Most machines can represent numbers in the range of 10 to the power of plus/minus some hundreds. For our algorithm, a rough estimate of the range of partition function values is done by the following argument: Assume the probability of a random microstate is $p = 2^{-N} \approx 10^{-N/3}$. For an N=$10^6$ sequence this implies a range of more than 300000 decades. Rescaling alone is therefore not sufficient and we must represent such extreme powers of 10 in software. Here we describe our method in the case of the fast version where W is calculated by recursion.

Our method is based on the rescaling briefly described by Yeramian (ref. 2) as a normalization performed every 50 bp's in the iteration. We choose a constant rescaling factor $F = 10^{-30}$ and a constant threshold $G = 10^{60}$ that should be some fraction of the machine limit. In the LR iteration, we do a rescaling for each step, where the summation of the total partition function surpasses the threshold, $Q_{\text{total}} > G$. All values in the vectors $U_{01}^{\text{LR}}$, $U_{11}^{\text{LR}}$ and $V_{10}^{\text{LR}}$ should be rescaled together with $Q_{\text{total}}$, but that would make the iteration run in time O(N$^2$). We only rescale the quantities $U_{01}^{\text{LR}}(i)$, $U_{11}^{\text{LR}}(i)$, $V_{10}^{\text{LR}}(i+1)$, $Q_{\text{total}}$, $W_i(m)$ for all m and $V_{10}^{\text{LR}}(1)$ in step i. The last two are rescaled to ensure that all subsequent values of the three vectors are also rescaled. To keep track of the vector values that were not rescaled, we define a "rescaling level" function L(i) that indicates the number of rescalings as a staircase function along the chain: $L(i) = j$ for $l_j \leq i < l_{j+1}$, where $l_1, l_2, ..., l_K$ are the LR iteration steps in which a rescaling is performed. Then the fully rescaled LR vectors are obtained as $V_{10}^{\text{LR}}(i+1)F^{K-L(i)}$, $U_{01}^{\text{LR}}(i)F^{K-L(i)}$ and $U_{11}^{\text{LR}}(i)F^{K-L(i)}$.

In the RL iteration we do rescalings in steps $r_1, r_2, ..., r_K$ given as $r_j = N + 2 - l_{K+1-j}$, that is, at chain units next to the units where a LR rescaling was performed. The quantities rescaled in step $r_j$ are $U_{01}^{\text{RL}}(r_j)$, $U_{11}^{\text{RL}}(r_j)$, $V_{10}^{\text{RL}}(r_j+1)$, $V_{10}^{\text{RL}}(1)$ and $W_{r_j}(m)$ for all m. The fully rescaled RL vectors are then obtained as $V_{10}^{\text{RL}}(i+1)F^{L(N+1-i)}$, $U_{01}^{\text{RL}}(i)F^{L(N+1-i)}$ and $U_{11}^{\text{RL}}(i)F^{L(N+1-i)}$.

We get a new set of equations for the various probabilities when inserting the expressions for the fully rescaled LR and RL vectors. We insert an extra factor of $F^K$ in the denominator of those ratios that have two subchain partition functions in the numerator, because both of them are rescaled K times. Equation 11 for the base-pairing probability $p_{\text{closed}}(i)$ is unchanged. The loop probabilities become

$$p_{\text{loop}}(a,b) = V_{10}^{\text{LR}}(a+1)\omega[2(b-a)]V_{10}^{\text{RL}}(N+2-b)F^{L(b)-L(a)} / \beta Q_{\text{total}}. \tag{16}$$

The tail probabilities become

$$p_{\text{right}}(i) = V_{10}^{\text{LR}}(i+1)F^{K-L(i)} / Q_{\text{total}} \tag{17}$$

and

$$p_{\text{left}}(i) = V_{10}^{\text{RL}}(N+2-i)F^{L(i)} / Q_{\text{total}}. \tag{18}$$

The helix probabilities become

$$p_{\text{helix}}(a,b) = U_{01}^{\text{LR}}(a)s^{\text{end}}(a)\left[\prod_{j=a+1}^{b} s^{11}(j)\right]s^{\text{end}}(b)U_{01}^{\text{RL}}(N+1-b)F^{L(b)-L(a)} / \beta Q_{\text{total}}. \tag{19}$$

## RESULTS AND DISCUSSION

The algorithm was implemented in Perl and different thermodynamic parameter sets taken from the literature have been used for calculating melting profiles for various human genomic sequences. In this article we will focus on a validation of the algorithm.

### Speed tests

Speed tests were performed to validate the algorithm. Figure 2 is a log-log plot of computation time [for computing a base-pairing probability profile $p_{closed}(i)$] versus sequence length for three different algorithms: a) the "fast" and b) the "slow" version of the algorithm described in this article (i.e. using a multiexponential approximation of the loop entropy factor and using the exact power function, respectively) and c) the "fast" algorithm described by Yeramian et al. (ref. 25) (using the same multiexponential approximation). The measured results are in accordance with the algorithmic time complexities listed in table 1. This confirms that: the algorithm described in this article has the same algorithmic time complexities as the PFF algorithms, that is, linear (a) for the multiexponential case and quadratic (b) for the exact case; and a speed-up is obtained with the "$LR \times RL$" method described in this article, as compared to the original algorithm of Yeramian et al., from quadratic (c) to linear (a) for the fast versions and from cubic (not tested) to quadratic (b) for the slow versions. By extrapolation in Figure 2, we find that a 1 million bp sequence would be processed in one minute with the linear (a) algorithm and in three weeks with the quadratic (b) algorithm.

### Numerical tests

Numerical tests were also performed to validate the algorithm. As a first test, we set all statistical weight factors equal to one, that is, all $s^{11}(i)$, $s^{010}(i)$, $s^{end}(i)$, $\omega[2(b-a)]$ and $\beta$. Then all microstates should have the statistical weight 1 and the total partition function should be equal to $2^N$. Indeed, we found the calculated $Q_{total}$ to be equal to $2^N$ for all chain lengths in the range $3 \leq N \leq 49$. This indicated that the algorithm takes each of the $2^N$ possible microstates correctly into account. For larger numbers of N, the precision is restricted by the floating-point format of the computer. For $N=10^6$ the rescaled total partition function was $Q_{total}=9.90065622930628 \times 10^{39}$ and rescaling was done $K=10033$ times, which means that the "true" total partition function is $Q_{total}/F^K=9.90065622930628 \times 10^{301029}$. Taking $\log_2$ of this number gives $N=1000000.0000000000015$. This high precision indicated that the rescaling scheme can handle ranges of thousands of decades in an accurate way.

Figure 3 shows the base-pairing probability profile $p_{closed}$ of a 4781 bp long sequence (GenBank accession number BC039060, an RB1-related cDNA sequence) calculated at T=84°C, at which temperature the average $p_{closed}(i)$ is 52%. The profile was calculated using the same three algorithms as in the speed test: The "approximation" (a) and the "exact" (b) version of the algorithm described in this article and the "approximation" (c) version described by Yeramian et al. (ref. 25). In order to do a controlled comparison with the algorithm of Yeramian et al., a thermodynamical model was chosen without explicit nearest neighbor effects, which can be simulated exactly by both algorithms. Parameters were taken from Fixman & Freire (ref. 22): $T_{AT}=342.5$ K, $T_{GC}=383.5$ K, and so forth. This simple model can be simulated here by setting $s^{010}(i) = \exp(\Delta(1 - T_i/T))$, $s^{end}(i) = \exp(\Delta(1 - T_i/T)/2)$ and $s^{11}(i) = \exp(\Delta(1 - (T_{i-1} + T_i)/2T))$. Data points for the three curves in the figure are on top of each other. The probabilities differ on average $6.58 \times 10^{-5}$ and maximally $8.6 \times 10^{-4}$ when calculated with the exact loop entropies (b) and with a ten exponentials approximation (a), which indicates that the multiexponential approximation is very good at this sequence length. The probabilities as calculated by this algorithm (a) and by Yeramian et al.'s algorithm (c) are identical within 10 decimals, indicating that the different algorithmic approaches to the same simple model do not introduce significant numerical differences.

## Loop Map

As an example of the other types of probabilities that can be calculated, Figure 4 is a loop map showing the probabilities and fluctuating positions of some of the most probable loops, helical regions and tails calculated under the same conditions as in Figure 3(a). Arcs above the axis illustrate loops and tails (open units) while arcs below the axis illustrate helical regions (closed units). Horizontal bars indicate the ranges of the fluctuations of the endpoints. For example, a loop is illustrated by an arc that connects two bars at intervals [$a_1,a_2$] and [$b_1,b_2$] and the indicated probability is for a loop to be bounded by 1's in these intervals, which is obtained as the sum

$$p_{\text{loop}}([a_1,a_2] \times [b_1,b_2]) = \sum_{i=a_1}^{a_2} \sum_{j=b_1}^{b_2} p_{\text{loop}}(i,j).$$

(20)

Within the [$a_1,a_2$]x[$b_1,b_2$] intervals, the loop probability attains a maximum at ($a_m,b_m$), and this is indicated in the loop map as the positions of the two arc ends. Note that fluctuations are not necessarily symmetrical in range to the left and right of the maximum point. The maximum probability itself, $p_{\text{loop}}(a_m,b_m)$, is not indicated since it is typically lower than 1%. Helical regions and tails are indicated similarly. The loop map shows a correspondence between neighboring open and closed regions. Note that many features of the probability profile in Figure 3 can be identified with the loops, tails and helical regions shown in Figure 4.

## Nearest Neighbor Thermodynamics

Why is nearest neighbor thermodynamics handled rigorously and generally using three types of statistical weight factors for nearest neighbor bp's, isolated bp's and helix-ending bp's? The following discussion is for the most common case of two possible bp types, AT and CG, although the algorithm can be applied to other cases as well (e.g., including mismatches).

Consider first a helical segment of more than one closed unit. As described in the previous section, a product of nearest neighbor factors and two helix-ending factors is assigned to the helical segment. The factor $s^{11}(i)$, describing the nearest-neighbor pair [i-1,i], takes on one of ten values depending on the type of dinucleotide. The factor $s^{end}(i)$, describing a helix-ending bp, takes on one of two values depending on the type of bp. A helical segment is thus described by a combination of 12 possible values. These values are obtained using thermodynamic parameters and depending on temperature, salt, and so forth.

Physically, there can in fact be four distinct helix-ending interactions, in contrast to our two possible values. If these four interactions were known, they could be incorporated in this algorithm by splitting $s^{end}(i)$ into separate factors for the left- and right-end of a helix, $s^{011}(i)$ and $s^{110}(i)$. Then 14 values would be used to describe helical segments. But as argued in the following, the combination of only 12 possible values can always generate the full information. In a quite general analysis, Gray (ref. 30, 31) has characterized the information that is needed for describing nearest neighbor additivity with constraints. He considers a region of bp's bounded by symmetrical "ends" that could be solvent or fixed sequence, and so forth. Using the concept of a fictitious end bp E/E', the base-paired region is written as [EX(1)...X(L)E']/ [EX'(L)...X'(1)E']. Both 5'-ends (E) are identical and both 3'-ends (E') are identical. His discussion applies to our internal helical segments as well, since we assume that loops and tail regions constitute symmetrical ends E/E'. Following Gray's analysis (for the case of two possible bp's), we conclude that: it is necessary to consider the end interactions properly (one initiation parameter $\sigma$ is not in general sufficient); although there are 14 different nearest neighbors (including end neighbors), at most 12 independent values can be uniquely derived from experiments; and combining 12 values is sufficient for an exact prediction of the nearest neighbor additive property. Such 12 values may not be physically interpretable as the actual local nearest neighbor contributions, instead we must think of them merely as model parameters needed for predictions.

Consider next an isolated bp. It is assigned the statistical weight factor $s^{010}(i)$, which takes on one of two values depending on the bp type. In Gray's analysis (ref. 30, 31) it is assumed that the description of the nearest neighbor property of a base-paired region applies to an isolated bp as well. With that assumption, the two possible values for $s^{010}(i)$ can be derived from the 12 values of $s^{11}(i)$ and $s^{end}(i)$. However, we allow for the assumption that isolated bp's have their own special (low) stabilities, which corresponds to having 3-body interactions, not just nearest neighbor, in an Ising-model context. This adds further generality to the algorithm, for example, some models totally exclude isolated bp's, which can be simulated here by setting $s^{010}(i)=0$. Otherwise, the two possible $s^{010}(i)$ values could be based on additional experimental data or perhaps from theoretical predictions of hydrogen bonding without stacking interactions.

The combination of the 12 helical values depends on their format. It is possible to translate from one format to another, typically by a linear transformation. Gray discusses two different formats: the independent short sequence (ISS) format and the individual nearest neighbor (INN) format. He argues that the ISS format provides a representation of the maximum amount and type of information that can be derived experimentally. However, he also shows that the INN format with ten NN values and two end (initiation) values, which is the format we use in our algorithm, is equivalent to 12 values in the ISS format, in the sense that they can give identical predictions. In addition, we note that 12 ISS values can be translated into our INN format. We therefore believe that our choice of format can represent the full information that is available from experiments, and that no other choice of format would give more accurate predictions. All these conclusions assumes nearest neighbor additivity.

The choice of an algorithm using the three types of statistical weight factors is therefore not made to advocate the 12+2 INN format for thermodynamic parameters. Rather, the point is generality. The belief is that any nearest neighbor parameter set can be included in a non-approximative way in the algorithm, by translating the parameters into the 12+2 INN format. As an example, we will indicate how parameters in two formats, the singlet format and the doublet format (ref. 7), can be translated. In the singlet format, the free energy change of a helical segment from a to b is a sum of b-a+1 bp contributions and b-a nearest neighbor "corrections",

$$\Delta G = \sum_{i=a}^{b} \Delta G_i^{bp} + \sum_{i=a+1}^{b} \Delta G_{i-1,i}^{NN}.$$

(21)

The statistical weight of the segment is

$$\exp(-\Delta G/RT) = s^{end}(a) s^{11}(a+1) s^{11}(a+2) \ldots s^{11}(b) s^{end}(b),$$

(22)

which is true if $s^{end}(i) = \exp(-\Delta G_i^{bp}/2RT)$ and $s^{11}(i) = \exp(-(\Delta G_{i-1}^{bp} + \Delta G_i^{bp} + 2\Delta G_{i-1,i}^{NN})/2RT)$. In the doublet format, the free energy change of the helical segment from a to b is a sum of b-a nearest neighbor terms,

$$\Delta G = \sum_{i=a+1}^{b} \Delta G_{i-1,i}^{NN}.$$

(23)

In this case, the statistical weight is obtained by $s^{end}(i) = 1$ and $s^{11}(i) = \exp(-\Delta G_{i-1,i}^{NN}/RT)$.

## *CONCLUSION*

The preferred choice of algorithm for computing Poland-Scheraga type of melting for specific DNA sequences of genomic length is one where computation time grows linearly with length, O(N), at least when used on a common PC. Most available linear programs (ref. 1, 23, 24) are implementations of the probability-based Poland-Fixman-Freire algorithms. Yeramian et al. introduced a partition function-based algorithm (ref. 25), but its computation time grows quadratically with length for the computation of a base-pairing probability profile. These algorithms introduce certain approximations in the handling of thermodynamic parameters for nearest neighbor stabilities. Based on the work of Yeramian et al., we propose an improved partition function-based algorithm where computation time grows linearly with length. The improvement is both with respect to the speed and the thermodynamic parameters. The speed-up is based on symmetrical recursions in the "left-to-right" and "right-to-left" directions along the chain. Nearest neighbor effects are included in a fully general and non-approximative way by using a format with three types of stabilities for nearest neighbor bp's, isolated bp's and helix-ending bp's.

As noted by Yeramian (ref. 25), a partition function algorithm, as opposed to a probability-based algorithm, has more generality and flexibility. The quantities involved in a partition function formalism are conveniently handled by standard calculational methods and more quantities than we have discussed in this article can be calculated. Commonly, the thermal ensembles of melting DNA have been represented by the base-pairing probability profiles $p_{closed}(i)$. In this article we have shown that in addition to $p_{closed}(i)$, the calculation of loop probabilities, $p_{loop}(a,b)$, tail probabilities, $p_{right}(i)$ and $p_{left}(i)$, and helix probabilities, $p_{helix}(a,b)$, is possible with our algorithm. In principle, these probabilities contain more information than the base-pairing probabilities alone. Analysis of these probabilities offers a direct identification of the two-state melting domains and their exact positions and sizes, as well as a characterization of regions where melting is not two-state. These matters are not clearly revealed by the base-pairing probabilities $p_{closed}(i)$ alone, but can be further investigated using the algorithm presented here.

# *TABLES*

| 1D Helix-Coil Models | | | 2D Secondary Structure Models | | |
|---|---|---|---|---|---|
| Algorithm | Time (approx/exact) | Memory (approx/exact) | Algorithm | Time | Memory |
| Poland-Fixman-Freire | $O(N)$ / $O(N^2)$ | $O(N)$ / $O(N)$ | McCaskill | $O(N^3)$ | $O(N^2)$ |
| Yeramian et al | $O(N^2)$ / $O(N^3)$ | $O(1)$ / $O(N)$ | Chen-Dill | $O(N^6)$ | $O(N^3)$ |
| Tøstesen et al | $O(N)$ / $O(N^2)$ | $O(N)$ / $O(N)$ | | | |

Table 1: Algorithmic complexities of various 1D and 2D approaches with regards to sequence length N. Time and memory requirements are shown for computing a base-pairing probability profile, which is a vector p(i) for the 1D helix-coil models and a matrix p(i,j) for the 2D secondary structure models. For the 1D helix-coil models, results are shown in the two cases of using a multiexponential approximation of the loop entropy factor and using the exact power function.

# FIGURES

Figure 1: Diagrammatic representation of the recursion relations for a) $U_{01}(i)$, b) $U_{11}(i)$ and c) $V_{10}(i+1)$.

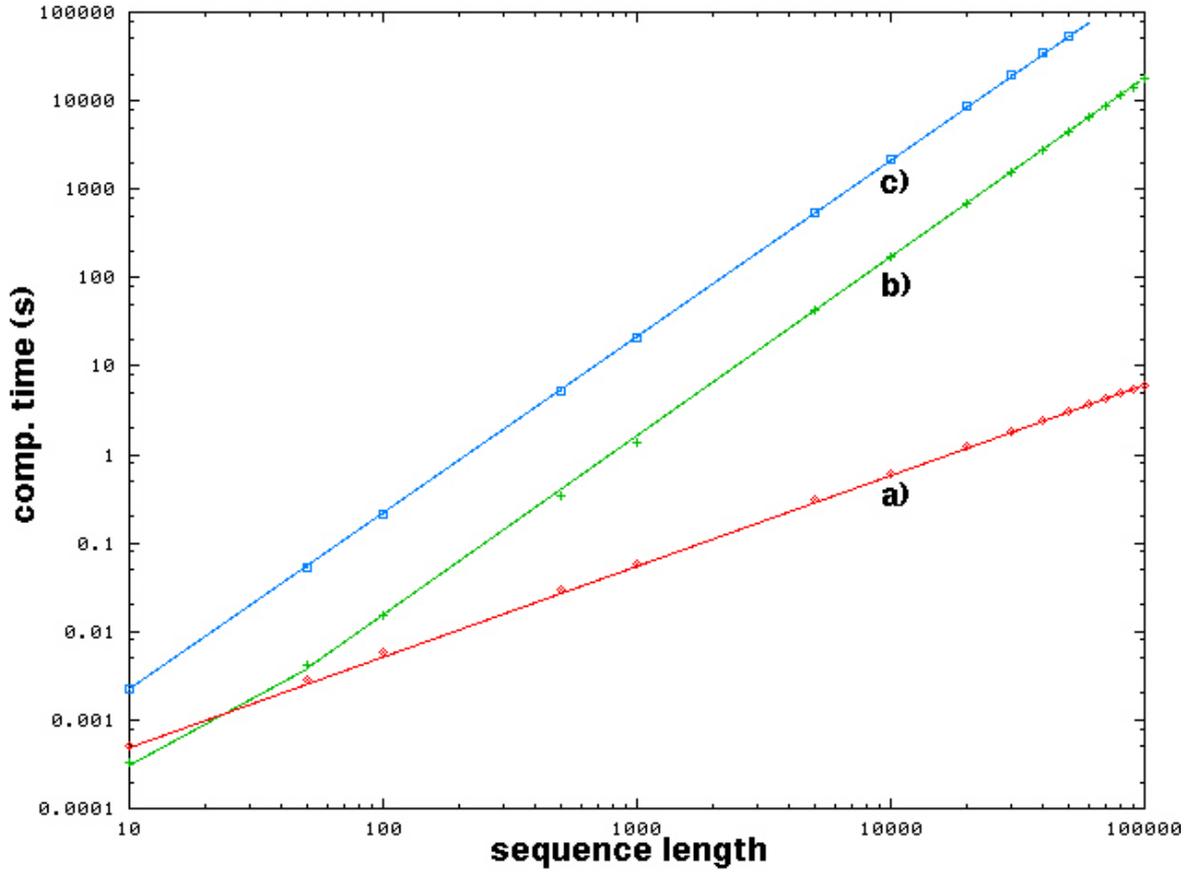

Figure 2: Log-log plot of average computation time versus sequence length N for three different algorithms: a) the "fast" version of the algorithm described in this article (using a multiexponential approximation of the loop entropy factor), which runs in time $O(N)$.  b) The "slow" version of the algorithm described in this article (using the exact power function for the loop entropy factor), which runs in time $O(N^2)$.  And c), the "fast" algorithm described by Yeramian et al. (ref. 25) (using the same multiexponential approximation), which runs in time $O(N^2)$. Times are in seconds for computing a base-pairing probability profile $p_{closed}(i)$. Algorithms were written in Perl and run on a 2.4 GHz PC.

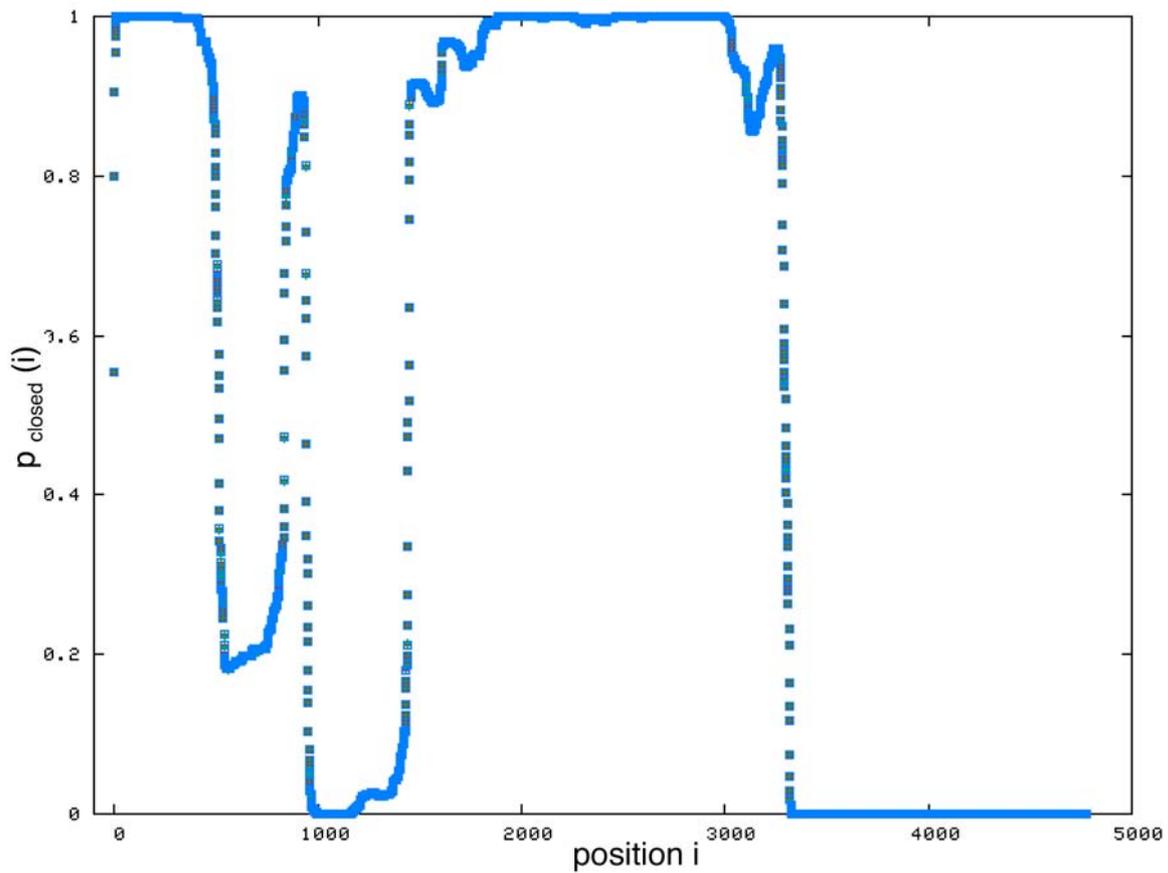

Figure 3: A probability profile $p_{closed}(i)$ of a 4781 bp sequence (GenBank accession number BC039060, an RB1-related cDNA sequence) calculated at T=84°C with the same three algorithms as in the speed test: The "approximation" (a) and the "exact" (b) version of the algorithm described in this article and the "approximation" (c) version described by Yeramian et al. (ref. 25). Data points for the three curves in the figure are on top of each other. Parameters were taken from Fixman & Freire (ref. 22) ) and $\alpha=1.8$ and $d=0$ were chosen.

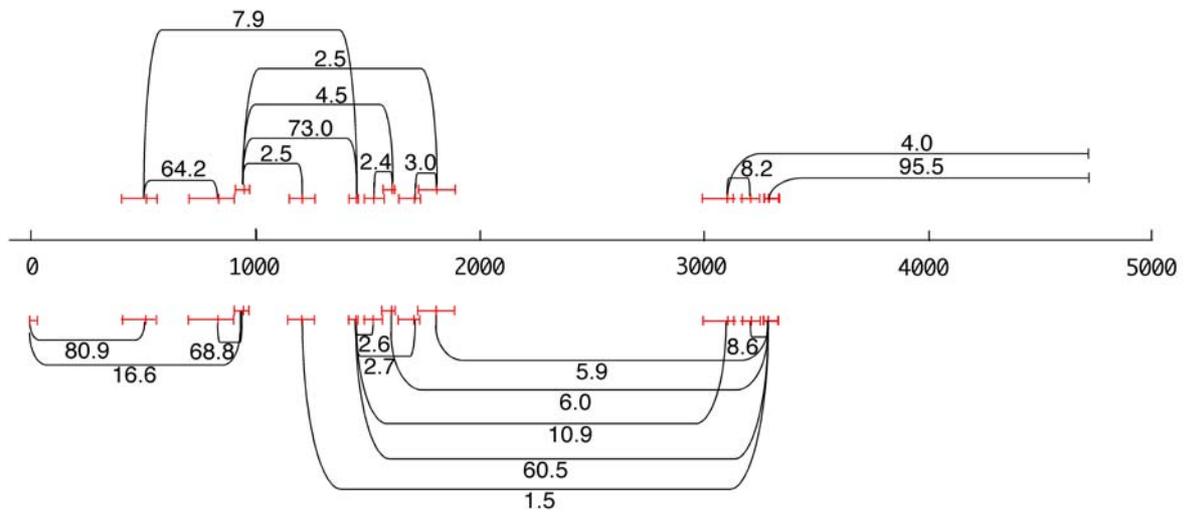

Figure 4: A loop map corresponding to Figure 3. The axis shows the position in the sequence. Arcs and "half arcs" above the axis indicate the positions of some of the most probable loops and tails (open unit regions). Arcs below the axis indicate the positions of some of the most probable helical regions (closed units). Fluctuations in these positions are indicated by horizontal bars at the ends of the arcs. Probabilities are summed over the bar intervals (see the text) and indicated for each arc in percent.